\documentstyle{article}

\newcommand{\be}{\begin{equation}}
\newcommand{\ee}{\end{equation}}
\newcommand{\bea}{\begin{eqnarray}}
\newcommand{\eea}{\end{eqnarray}}
\newcommand{\nn}{\nonumber}

\newcommand{\p}[1]{(\ref{1})}
\def\theequation{\arabic{section}.\arabic{equation}}
\begin{document}
\thispagestyle{empty}
\rightline{hep-th/9510050}
\vglue 2.0cm

\begin{center}{\Large\bf
 Practical scheme of reduction to gauge invariant variables}\\
\vspace*{10mm}
{
\bf{ G.\ Chechelashvili ${}^{\ast}$}, G.\ Jorjadze ${}^{\ast,}$
\footnote{Electronic address: jorj@imath.kheta.ge}
and N.\  Kiknadze ${}^{\dagger}$}\\
\vspace*{1cm}
{\it ${}^{\ast}$ Tbilisi Mathematical Institute,\\
 380093 Tbilisi, Georgia\\
\vspace*{0.5cm}
${}^{\dagger}$ Tbilisi State University}

\vspace*{2cm}
\end{center}

\begin{abstract}
{For systems with first class constraints the reduction
scheme to the gauge invariant variables is considered.
The method is based on the analysis of restricted 1-forms
in gauge invariant variables.
 This scheme is applied to the models of
 electrodynamics and Yang-Mills theory. For the
 finite dimensional model with $SU(2)$ gauge group of symmetry the
possible mechanism of confinement is obtained.} \end{abstract}
\vspace*{5mm}

\newpage
\setcounter{footnote}{0}

\section{Introduction} \vspace*{0.5cm}

Most of interesting physical models and theories are described by the
gauge invariant Lagrangians, which are singular and in Hamiltonian formulation
lead to the constrained systems [1]. For the constrained Hamiltonian systems
there are in principle two ways\footnote{In this paper we do not
consider the path integral approach} of quantization [1-2]:

1. ``First quantize and then reduce".

2. ``First reduce and then quantize".

The present paper deals mainly with the reduction procedure of the
latter. For gauge theories this procedure is a restriction on the
constraint surface $\cal M$ and then farther reduction to the
physical phase space $\tilde {\cal M} = {\cal M}/G$, which is the
space of gauge orbits.

If the action of the gauge group (G) on the constraint surface  ($\cal M$) is
regular, then the manifold of orbits ($\tilde {\cal M} = {\cal M}/G$ ) is
well defined and it possesses the symplectic structure. Coordinates
on $\tilde{\cal M}$ are gauge invariant, true physical degrees of
freedom.

Quite often, in practical applications this theoretical scheme of reduction
encounters technical problems related to the explicit construction
of $\tilde{\cal M}$ with its symplectic structure: apart from the
mathematical difficulties, the physical content of the true degrees
of freedom may be quite unpredictable.

Widely used practical reduction scheme is a gauge fixing procedure
by some conditions $\chi (p,q)=0$ [1],[3].  For simple cases the
explicit form of the true physical variables is obvious and this
reduction scheme works perfectly.  But in general, as it was shown
in [4] (namely for the Yang-Mills theory), the space of gauge orbits
(${\cal M}/G$) cannot be obtained  by ``simple" gauge fixing. Problem of
gauge fixing, of course, reflects the above mentioned possible
non-trivial structure of a physical phase space [1],[5].

Another reduction scheme can be based on the gauge invariant
variables ({\bf GIV}s) [1-2]. As a rule, the {\bf GIV} is constructed
from the structure of gauge transformations. If one can find the
complete set of {\bf GIV}s, then it allows to describe
the physical phase space ${\cal M}/G$ with its symplectic
structure. This paper deals with such gauge invariant approach
by using of restricted 1-forms. We also consider situations when
only a part of {\bf GIV}s is known.  Analysis of a structure
of restricted 1-form helps to find the remaining part of {\bf GIV}s.

Note that the reduction scheme with 1-forms for arbitrary constrained
systems was proposed in [6]. In these papers elimination of extra
variables was based on the Darboux theorem. Sometimes
Darboux theorem is not effective in applications and choice of {\bf
GIV}s is just a practical way for the realization of this reduction
program for gauge theories.

The paper is organized as follows: In Section 2 the reduction
scheme to the {\bf GIV}s is introduced and, for the illustration,
simple examples are considered. One more example of $(2+1)$-dimensional
massive photodynamics is given in Appendix.  In Section 3
this scheme is applied to the finite dimensional system with $SU(2)$
gauge group of symmetry.  This system can be considered as a toy
model of the Yang-Mills theory with fermions. It is shown that there is
an essential difference between this $SU(2)$ and the corresponding
$U(1)$ model. Structure of {\bf GIV}s in $SU(2)$ case can be
interpreted as the confinement phenomenon. In Section 4 we study
infinite dimensional model, where gauge group is any semi-simple
one. The {\bf GIV}s are constructed and full reduction is accomplished.
It is shown that the model is equivalent to the Yang-Mills theory
with some boundary conditions. Final section is for remarks and
conclusions.

\vspace*{1cm}

\section{ Reduction scheme in gauge invariant variables}
\vspace*{0.5cm}

 Starting from the gauge invariant Lagrangian $L = L(q_k,\dot
{q_k}) \quad (k=1,...,N)$ and using the Dirac's procedure [1a], or
the first order formalism [6] we arrive to the extended phase space
$\Gamma $ with coordinates $(p_k ,q_k )$ and the action
\bea
S=\int p_k dq_k -[ H(p,q) + \lambda _a \phi _a (p,q)]
            dt, \\
\nn k=1,...,N;~~~a=1,...,M;~~(N > M),
            \eea 
where $\phi_a (p,q)$ are constraints,
$H(p,q)$ is a canonical Hamiltonian and $\lambda_a$ are
Lagrange multipliers.  The constraint surface --- $\cal M$ is
            defined by
\be \phi _a (p,q) = 0 \ee 
and the
following relations are fulfilled:
\be \{ H,\phi_a \}_{\Gamma} =
d_a^b\phi_b,~~~~~~\{ \phi_a, \phi_b \}_{\Gamma} = f_{ab}^c
 \phi_c.
\ee 
 Index $\Gamma$ on the left hand side
indicates that Poisson brackets are calculated on the extended
phase space.

Function $\xi = \xi (p,q)$ is called {\bf GIV} [1]
if $\xi |_{\cal M} \not= 0 $ and
\be
 \{ \xi ,\phi_a \} _{\Gamma } = \tilde {d}_a^b\phi_b,
\ee
where $|_{\cal M}$ denotes restriction on ${\cal M}$ and functions
$\tilde {d}_a^b$ (as  well as $d_a^b$ and $f_{ab}^c$ in (2.3)) are
assumed to be regular in the neighbourhood of $\cal M$.

Each {\bf GIV} --- $\xi $  has the class $\{\xi \}$ of equivalent
{\bf GIV}s on $\Gamma $ [1]. A gauge invariant function $\tilde \xi $ is
equivalent to $\xi $ if $\tilde {\xi}|_{\cal M}$ = $\xi |_{\cal M}$ .
On the other hand, the function
$\xi |_{\cal M}$ is a constant along the gauge orbit (on ($\cal M$))
and it defines the function $\bar \xi $ on the physical space
$\tilde {\cal M} = {\cal M}/G$. Thus $\{\xi \}$, $\xi |_{\cal M}$
and $\bar \xi $ denote the {\bf GIV} --- $\xi $ in different context.
If there is no ambiguity, we will use the notation $\xi$ for all of them.

Maximal  number of {\bf GIV}s (2.4), which are functionally independent
on the constraint surface $\cal M$,  is $2(N-M)$ [1b]. Suppose that
$\{\xi^\alpha : \alpha = 1,..., 2(N-M)\}$ is such
complete set of {\bf GIV}s. Then one can prove [6] that
\bea
&1.&~~ p_k dq_k |_{\cal M} = \theta_1 + \theta_2 ,~~~ \mbox{with} \nn \\
&a)&~~ d\theta_1 = 0, \nn \\
&b)&~~ \theta_2 = \theta_\alpha (\xi ) d\xi^\alpha,   \\
&c)&~~ det\omega_{\alpha\beta}\not= 0,~~~ \mbox{where}~~~~
\omega_{\alpha\beta}(\xi ) = \partial_{\alpha }\theta_{\beta} -
\partial_{\beta }\theta_{\alpha}; \nn \\
&2.&~~ H(p,q) |_{\cal M} = h(\xi ). \nn
\eea
Main statement of (2.5) is that after restriction on the constraint surface
$\cal M$, dependence on extra (non-physical) variables is present
only in the term
$\theta_1$, which is a ``total derivative".

Since $d\theta_1 = 0 $, it gives no contribution to the variation of
a restricted action. We can neglect it and for the
reduced system  get
\be
S|_{\cal M} \equiv \tilde S = \int \theta_{\alpha} (\xi )d\xi^{\alpha} -
h(\xi )dt.
\ee
 Hence  dynamics for {\bf GIV}s is described by the Hamilton equations
\be
\dot {\xi^\alpha} = \omega^{\alpha \beta}(\xi ) \partial_{\beta} h(\xi),
\ee
where $ \omega^{\alpha \beta}(\xi ) $ is the inverse to the symplectic matrix
$ \omega_{\alpha \beta} = \partial_{\alpha }\theta_{\beta} -
\partial_{\beta }\theta_{\alpha}$ and it defines the Poisson brackets of the
reduced system
\be
\{ \xi^{\alpha},\xi^{\beta}\}_{\tilde {\cal M}} = \omega^{\alpha \beta} (\xi ).
\ee
So the reduced system (2.6)-(2.8) is an ordinary (non-constrained)
Hamiltonian system which can be quantized.

It should be noticed that, in general, any $2(N-M)$ number of {\bf GIV}s
are only local
coordinates on the physical phase space ${\tilde {\cal M}}$ and respectively
(2.5)-(2.8) have the local meaning. Global description can be achieved by the
set of {\bf GIV}s which defines the global structure of physical phase space
${\tilde {\cal M}}$. Number of such {\bf GIV}s is greater than $2(N-M)$,
but on the
constraint surface there are relations among them. Just these relations
define the geometry of
${\tilde {\cal M}}$. For the illustration let us consider the following
example of (2.1)-(2.3) [7a]:
\be
S = \int\vec p\cdot d\vec q - [\lambda_1 \phi_1 + \lambda_2 \phi_2 ] dt.
\ee
Here $\vec p$ and $\vec q$ are vectors of ${\bf R}^3$, canonical Hamiltonian is
zero,
$$\phi_1 = \vec p\cdot\vec q ,~~~~ \phi_2 ={\vec p}\,{}^2 {\vec q}\,{}^2 -
( \vec p\cdot\vec q )^2 -r^2
$$
and $r$ is a parameter. These constraints are Abelian ($\{ \phi_1 , \phi_2 \}
=0$) and the second constraint $\phi_2 $ can be written in the form
$$
\phi_2 = {\vec J}\,{}^2 - r^2
$$
where $\vec J =  \vec q \times \vec p $ is an angular momentum.

It is clear that the physical phase space is two dimensional and the components
of angular momentum $\vec J $ are {\bf GIV}s (they commute with constraints,
since constraints
are $O(3)$ scalars). On the constraint surface these three components are
related by
$\vec J \cdot \vec J = r^2 $ and define the physical phase space
$\tilde {\cal M} $
as the two dimensional sphere. So any two coordinates (as well as the 1-forms
$\theta_1$ and $\theta_2$) are only local ones (on the phase space geometry
of constrained systems see [5a]).

Described reduction scheme ((2.5)-(2.8)) can be used when all $2(N-M)$
{\bf GIV}s are known.
For the practical application of the scheme one can introduce any
additional (to {\bf GIV}s) variables $\eta^1,...,\eta^M$ to
have the coordinate system $$(\xi^1,...,\xi^{2(N-M)},
\eta^1,...,\eta^M)$$ on ${\cal M}$. Then calculating restricted
1-form $p_k dq_k |_{\cal M}$ in these coordinates and taking its
differential we can find the symplectic form
$\omega=\omega_{\alpha\beta}(\xi )d\xi^\alpha \wedge d\xi^\beta$.
Note that in practical calculations it is possible to single out the
1-form $\theta_2=\theta_\alpha(\xi)d\xi^\alpha$ and arrive to (2.6).

Application of this procedure to the model (2.9) gives
$\theta_2 = z d\varphi$, where $z$ and $\varphi$ are the
cylindrical coordinates on the sphere:
$$
J_1=\sqrt{r^2-z^2}cos\varphi, \quad
J_2=\sqrt{r^2-z^2}sin\varphi, \quad
J_3=z.
$$
It is clear that $zd\varphi$ is not the global 1-form, but
its differential has a continuation to the well-defined
symplectic form on the sphere [8]
$$
\omega = -\frac {J_1(dJ_2\wedge dJ_3)+J_2(dJ_3\wedge dJ_1)+
J_3(dJ_1\wedge dJ_2)}{r^2}.
$$
After this the system can be
quantized by the geometric quantization [9] (see also [7a],[10b,c]).
Consistent quantum theory exists only for the discrete values of
the parameter $r$.

Generalization of the scheme to the infinite dimensional case is
straightforward (in Appendix we present the example of massive
photodynamics in $(2+1)$ dimensions). If we use the Dirac's observables [11]:
\be
\psi_{in}= e^{i\Delta^{-1}\left(\vec\nabla\vec A\right)}\psi
\ee
in the ordinary QED, we will easily obtain photons in the Coulomb gauge and
the ``four-fermion interaction" for the ``dressed fermions" (compare to
the example in Section 3 and see [6], [12]).

Note that the commutation relations of the complete set of {\bf GIV}s (2.8)
can be derived by calculations of Poisson brackets on the extended phase space
too [1]. This more standard procedure is based on the fact that the Poisson
brackets of any two {\bf GIV}s is again {\bf GIV}. Indicated procedure
and the scheme
described in this paper ((2.5)-(2.8)) are almost equivalent. Only, sometimes,
calculations of differential forms is more practical
(especially, when the canonical quantization is not applicable [9]).

In general, from the structure of gauge transformations one can easily
find only some part of {\bf GIV}s and  construction
of the complete set (2.5) is troublesome.
In many cases, our approach with differential forms,  can be effectively
used for the solution of this problem too.

 Let us consider situation when we know the set of {\bf GIV}s
$\{\xi^\mu : \mu =1,...,K\}$, where $ N-M \leq k \leq 2(N-M) $. We can
introduce any additional variables $\eta^1,...,\eta^{2N-M-K}$ to have the
coordinate system on ${\cal M}$ and
calculate the restricted 1-form $p_k dq_k|_{\cal M}$. Suppose that we
can single out ``total derivatives" and the differentials $d\xi^\mu $
in the form
\be
p_k dq_k|_{\cal M} = dF(\xi ,\eta ) + \theta_\mu(\xi ,\eta ) d\xi^\mu .
\ee
Then using (2.5) we can easily conclude that $\theta_\mu(\xi ,\eta )$
will be {\bf GIV}s. Note that passing to the {\bf GIV}s --- $\xi^{\mu }$
helps to get the form (2.11).
For the illustration of this method we  apply it to the
relativistic particle [1c], with the 1-form
$\theta =\vec p d\vec q - p_0 dq_0$
and the constraint surface $\cal M$: $p^2 - m^2 = 0, (p_0 >0)$.
The momenta $\vec p$ are gauge invariant and after restriction
on $\cal M$ we have
$$
\theta |_{\cal M}=\vec p d\vec q - \sqrt{{\vec p}\,{}^2 + m^2} dq_0.
$$
One can easily rewrite it in the following form
$$
\theta |_{\cal M} = d(\vec p \cdot \vec q -
\sqrt{{\vec p}\,{}^2 + m^2} q_0)
-(\vec q - \frac{\vec p}{\sqrt{{\vec p}\,{}^2 +
m^2}}q_0 )d\vec p.
$$
Evidently the coefficients of the differentials $-d\vec p$
are {\bf GIV}s, canonically conjugate to $\vec p$:
$$
     \vec Q = \vec q - \frac{\vec p }{\sqrt{{\vec p}\,{}^2 + m^2}}q_0 .
$$

Gauge invariance of $\vec Q$ also can be established from the relation
 \be
 \vec L = \sqrt{{\vec p}\,{}^2 + m^2} \vec Q ,
 \ee
where $\vec L$ are the generators of Lorentz transformations. Since  all
generators of the Poincare group $(P_{\mu}, M_{\mu\nu})$ are {\bf GIV}s,
 the same property have the coordinates $\vec Q$. On the constraint surface
$p^2 - m^2 = 0, (p_0 >0)$ all these are the functions only
of the reduced variables $ \vec p$  and  $\vec Q$.

Reduced system  can be easily quantized in momentum representation:
$\hat {\vec p} = \vec p$ and $ \hat {\vec Q} = i\hbar \vec \nabla$. Operator
ordering problem arises only for the generators (2.12). Then the standard
Lorentz covariant measure of a scalar product
$$
<\Psi_2 |\Psi_1> = \int \frac{d^3\vec p}{\sqrt{{\vec p}\,{}^2 + m^2}}
\bar {\Psi_2}(\vec p) \Psi_1 (\vec p)
$$
corresponds to the ordering $\hat {\vec L} =
i\hbar \sqrt{{\vec p}\,{}^2 + m^2} \vec \nabla$.

\vspace*{1cm}

\setcounter{equation}{0}
\section{The finite dimensional models with $U(1)$ and $SU(2)$ gauge
symmetries}

\vspace*{0.5cm}

In this section we consider the finite dimensional model with $SU(2)$ gauge
group of symmetry. From the beginning it is difficult to see all
{\bf GIV}s and
we use the method described at the end of Section 2. Obtained structure
of {\bf GIV}s is quite unexpected. For comparison we present the corresponding
$U(1)$ model too. These $U(1)$ and $SU(2)$ models can be  considered as the
toy models of the electrodynamics and the Yang-Mills theory (with matter),
respectively.
In classical description all ``fields" are assumed to be $c$-numbers.

\vspace*{0.7cm}

{\bf A. The model with $U(1)$ symmetry}

\vspace*{0.5cm}

Let us consider the action
\be
S=\int {dt [\frac{i}{2}(\bar \psi \dot \psi - \dot {\bar \psi} \psi) -
m\bar \psi \psi + A_0 (\bar{\psi}\psi -k E) + E \dot A
-\frac{1}{2} E^2]},
\ee
where all ``fields" ($\bar \psi , \psi , A_0, A, E$) are functions only
of the time $t$; $m$ and $k$ ($k \neq 0$) are parameters.
Similarity to the electrodynamics is apparent from the notations.
At the same time (3.1) has the form (2.1),
where $A_0\equiv\lambda (t)$ is a Lagrange multiplier and $\phi
\equiv\bar{\psi}\psi - k E $
is a constraint (we use the time derivatives instead of differential
form
when it is convenient).

Non-zero Poisson brackets are
$$
\{\psi ,\bar \psi\} = i,~~~~~~~\{E , A\} = 1
$$
and we get the gauge transformations
\bea
&\psi(t)& \longrightarrow e^{+i\alpha (t)}\psi (t), \qquad
\bar \psi(t)\longrightarrow e^{-i\alpha(t)}\bar \psi (t), \nn \\
&A(t)& \longrightarrow A(t) + k\alpha (t), \qquad
E(t) \longrightarrow E(t).
\eea
Then
$$
A_0 (t) \longrightarrow A_0 (t) + \dot{\alpha}(t)
$$
leaves the action (3.1) invariant.

Reduced system is two dimensional and two {\bf GIV}s can be chosen as
\be
\Psi_{\mbox {inv}} = e^{\frac{-i}{k}A}\psi , ~~~~
\bar \Psi_{\mbox {inv}} = e^{\frac{i}{k}A}\bar \psi
\ee
(compare to (2.10)). Here, the reduction procedure (2.5) is trivial and we get
\be
\tilde S=\int {dt \left[\frac{i}{2}(\bar {\Psi}_{\mbox {inv}}
\dot {\Psi}_{\mbox {inv}}
 - \dot {\bar \Psi}_{\mbox {inv}} \Psi_{\mbox {inv}}) -
m\bar {\Psi}_{\mbox {inv}} \Psi_{\mbox {inv}} -
\frac {1}{k^2}({\bar {\Psi}_{\mbox {inv}} \Psi_{\mbox {inv}}})^2\right]}.
\ee
So the ``gauge field" $A$ has vanished and
physical excitations are only the ``dressed fields" $\Psi_{\mbox {inv}}$
(with "four-fermion interaction").

This model has a simple generalization in case of multi-component gauge field
$\vec A$ with gauge transformations
$$\vec A \longrightarrow \vec A
+\vec k\alpha ,
$$
where $\vec k$ are parameters (${\vec k}\,{}^2 \neq 0$).
The {\bf GIV} $\Psi_{\mbox {inv}}$ is constructed similarly to (3.3)
(or (2.10)).
Then, after reduction, ``longitudinal" (to the $\vec k$) component of the
gauge field $\vec A$
vanishes and physical variables are only ``transverse"  ones and
the constructed ``dressed field" $\Psi_{\mbox {inv}}$.
So, for these Abelian models, the structure of {\bf GIV}s is very similar
to the physical observables of the electrodynamics [6],[12].

\vspace*{1cm}

{\bf B. The model with $SU(2)$ symmetry}

\vspace*{1cm}

For the model with $SU(2)$ gauge group of symmetry we consider the action
\be
S=\int {dt\left[\frac{i}{2}(\bar \psi_{\alpha} \dot \psi_{\alpha} -
\dot {\bar \psi_{\alpha}} \psi_{\alpha}) -
m\bar \psi_{\alpha} \psi_{\alpha} + \vec A_0 (\vec j + \vec J )
+ \vec E \dot {\vec A}-\frac{1}{2}{\vec E}^2\right]}.
\ee
Here $\psi_{\alpha}$ are 2-component spinors ($\alpha =1,2$), $m$ is a
parameter, $\vec A$ and $\vec E$ are three dimensional vectors,
$\vec A_0 $ --- lagrange multipliers and the angular momenta $\vec j$
and $\vec J$ are given by
\be
   \vec j = \bar \psi \frac{\vec \sigma}{2}\psi,~~~~~~\vec J =
\vec A \times \vec E ,
\ee
where $\vec \sigma $ are the standard Pauli matrixes.

Connection with the Yang-Mills theory is obvious.

Non-zero Poisson brackets are
\be
\{\psi_{\alpha} ,\bar \psi_{\beta}\} = i \delta_{\alpha\beta} ,
{}~~~~~~\{E_m , A_n\} = \delta_{mn},~~~(m,n) =1,2,3
\ee
and the constraints $\vec \phi =\vec j - \vec J$ generate the
gauge transformations:
$$
\psi\to\omega\psi,              \quad
\bar\psi\to\bar\psi\omega^{-1}, \quad
{\bf A}\to\omega {\bf A} \omega^{-1},
\quad {\bf E}\to\omega {\bf E}\omega^{-1},
$$
where $\omega (t) \in SU(2)$ and
\be
{\bf A}\equiv \frac {1}{2}\vec A\vec \sigma,
 \qquad
{\bf E}\equiv \frac {1}{2}\vec E\vec \sigma.
\ee
 Then for $ {\bf A}_0\equiv \frac {1}{2}\vec A_0\vec \sigma$ we get
$
{\bf A}_0\to\omega{\bf A}_0\omega^{-1} -i\dot \omega\omega^{-1}.
$

 Any scalar product of the vectors $\vec A, \vec E, \vec J, \vec j  $
will  be {\bf GIV}. But on the constraint surface ($\vec j + \vec J = 0$) only
three of them are functionally independent.

If we choose these independent {\bf GIV}s as:
\be
l_0=\frac{1}{4}(\vec A\,{}^2+\vec E\,{}^2), \quad
l_1=\frac{1}{2}(\vec E\vec A), \quad
l_2=\frac{1}{4}
(\vec A\,{}^2-\vec E\,{}^2),
\ee
then from (3.7) we get the $SL(2,{\bf R})$ algebra:
\be
\{l_\mu,l_\nu\}=\epsilon_{\mu\nu\sigma}g^{\sigma\rho}l_\rho,
\quad \mbox{where} \quad
g^{\mu\nu}=diag(+,-,-,).
\ee

 Since there are three constraints,
 the physical phase space is 4-dimensional. To construct the fourth {\bf GIV}
and find the full symplectic structure we use the method of Section 2
(see (2.11)).

For the parameterization of the constraint surface we introduce new variables
$(j, \Phi ; h, \phi)$:
\be
j=\frac{1}{2}(h_1+h_2), \qquad
h=\frac{1}{2}(h_1-h_2), $$$$
\Phi=\varphi_1+\varphi_2, \qquad
\varphi=\varphi_1-\varphi_2,
\ee
where
$$
\psi_\alpha=\sqrt{h_\alpha}e^{-i\varphi_\alpha}, \quad
\bar\psi_\alpha=\sqrt{h_\alpha}e^{i\varphi_\alpha} \quad (\alpha =1,2).
$$
Then for the 1-form we get
\be
\frac{i}{2}(\bar\psi_{\alpha} d \psi_{\alpha} -
\psi_{\alpha}d {\bar\psi_{\alpha}} ) = jd\Phi+hd\varphi .
\ee
The vector $\vec j$ (3.6) in these new coordinates will take the form
$$
\vec j = \left(\begin{array}{c}
\sqrt{j^2-h^2}cos\varphi \\
\sqrt{j^2-h^2}sin\varphi \\
h
\end{array}\right)
$$
and $\vec j\,{}^2= j^2$.
Note that on the constraint surface we have (see (3.9)):
$l^\mu l_\mu= {j^2}/4 $
and for the fixed $ j$ the commutation relations (3.10) define well
known symplectic structure on this hyperboloid (see e.g. [13]).

 If we introduce the ortho-normal basis ($
\vec e_i\cdot\vec e_k=\delta_{ik}, ~~~ \vec e_i\times\vec e_j
= \epsilon_{ijk}\vec e_k
$ ) :
$$
\vec e_1 = \left(\begin{array}{c}
-sin\varphi \\
cos\varphi \\
0
\end{array}\right), \qquad
\vec e_2 = -\frac{h}{j}\left(\begin{array}{c}
cos\varphi \\
sin\varphi \\
-\frac{\sqrt{j^2-h^2}}{h}
\end{array}\right), \qquad \vec e_3=\frac{\vec j}{j},
$$

then  $\vec A$ and $\vec E$ can be parameterized as follows:
$$
\vec A=\vec e_1q_1+\vec e_2q_2,~~~~~~\vec E=\vec e_1p_1+\vec e_2p_2,
$$
where
\be
p_1q_2-p_2q_1 = j.
\ee
Calculating the restricted 1-form $\vec E d\vec A|_{\cal M} $ in these new
coordinates and using (3.13)  we obtain
\be
\vec E d\vec A|_{\cal M} =p_1dq_1+p_2dq_2 -hd\varphi .
\ee
Comparing (3.12) and (3.14), we see that there is a cancellation of the 1-form
$hd\varphi$. This means that the corresponding degrees of freedom vanish.

Now, it is convenient to introduce the polar coordinates
for two-vectors $(q_1,q_2)$ and $(p_1,p_2)$:
$$
q_1=rcos\beta,~~~~~~~~~~~p_1=\rho cos\gamma,
$$
$$
q_2=rsin\beta,~~~~~~~~~~~p_2=\rho sin\gamma.
$$
Then three of them ($r, \rho$ and $(\beta -\gamma)$) are connected
with the {\bf GIV}s (3.9):
$$
r^2=2(l_0+l_2)\equiv l_+, ~~~~~~\rho^2 = 2(l_0-l_2)\equiv l_-,
{}~~~~~~r\rho cos(\beta -\gamma )=2l_1.
$$
 Using these relations we finally get the reduced 1-form
\be
\theta|_{\cal M}=jd\vartheta +l_1\frac {dl_+}{l_+}~~~~~
 \mbox {where}~~~~ \vartheta=\Phi-\beta.
\ee
So the coordinate $\vartheta=\Phi-\beta $ is the fourth {\bf GIV}.
Respectively the reduced Hamiltonian takes the form
\be
H|_{\cal M}=2mj+\frac {j^2 + 4l_1^2}{l_+}
\ee
and this is the complete reduction.

Note that the second part of the reduced 1-form
$ l_1 {d{(\ln l_+ )}}$
defines the above mentioned symplectic structure on the hyperboloid
$l^\mu l_\mu=\frac{1}{4}j^2$ [13].

We see that the physical picture of this
reduced system essentially differs from the corresponding Abelian case.
Here, after reduction some part of degrees of freedom
of the ``gauge field" $(A)$, as well as some
part of ``matter field" ($\psi $) degrees of freedom have vanished.
Below we shall see that in
quantum theory vanishing of ``matter field" degrees of freedom can be
interpreted as the confinement phenomenon.

Geometric quantization [9] is a natural way for the construction of quantum
theory of the reduced system (3.15-3.16), but in principle one can use
canonical quantization too. For this aim it is convenient to introduce (global)
``creation" and ``annihilation" variables
 \be
{ a^+}=\sqrt{j}e^{i\vartheta}, \qquad
{ a}=\sqrt{j}e^{-i\vartheta}
\ee
and in quantum theory we get the discrete values of
$j = a^+ a$. Then, quantization of the system with  the canonical 1-form
$l_1 {d{(\ln l_+ )}}$
and the Hamiltonian (3.16) (for obtained discrete values of $j$), gives
the irreducible representations of $SL(2,{\bf R})$ group (see e.g. [13a]).

Next, from (3.11) we have the relation
$N\equiv\bar \psi_\alpha \psi_\alpha = 2j $.
It is natural to interpret the corresponding operator ($\hat {\bf N}\equiv
 2\hat {\bf j}$) as the $\psi $ particle number operator.
In quantum theory we have
$$ [\hat {\bf N},\hat {\bf a^+}]=2\hat {\bf a^+}
$$
where $\hat {\bf a^+}$ is a physical creation operator (3.17). So in physical
excitations (created by the operator $\hat {\bf a^+}$) there are states only
with even number of ``fermions". This fact also can be seen from the structure
of the variable $a^+$ (see (3.17) and (3.11)). It has the phase factor
$e^{i(\varphi_1+\varphi_2)}$. So in quantum case it will create (see [14])
the pairs of ``dressed" $\psi $-particles.

Note that for the similar finite dimensional constrained systems such
``confinement"-like phenomenon has been derived by the
``first quantize and then reduce" method (see [15]). In that approach
reduction of the extended ``Hilbert" space by the conditions
$\hat\phi_a|\Psi_{phys}>=0$ forbids the states with certain quantum numbers.

\vspace*{1cm}
\setcounter{equation}{0}

\section{Field theory models with non-Abelian gauge group of symmetries}
\vspace*{1cm}

For the finite dimensional models of the previous section the gauge group $G$
acts on the configuration space of ``gauge field" ($A$) and on the phase
space of ``matter field" ($\psi $). This is the standard situation for
Yang-Mills theories.

Using notations (3.8) we have
\be
\vec E d\vec A = <{\bf E}, d{\bf A}>,
\ee
where $<~,~>$ is a scalar product in corresponding Lie algebra $\cal A$.
Thus, the Lie algebra $\cal A$ can be interpreted as the configuration
space of a ``gauge field" $\vec A$ and trivial cotangent bundle as the phase
space.

If one takes a manifold of semi-simple Lie group ($G$)  as the
configuration space, then there are the natural actions (left and right) of
$G$ on this manifold and  one can construct similarly the  gauge theory
where phase space is the cotangent bundle [16]
$
T^*G = \{(g,R)| g\in G, R\in {\cal A}\}.
$
On $T^*G $ the symplectic form is given by
\be
\omega = d\theta ,~~~ \mbox {with}~~~ \theta = <R, g^{-1}dg>.
\ee
Generators of the left and right transformations $(g \longrightarrow
\omega g, ~  g \longrightarrow g \omega) $
are respectively left and right currents $ (L\equiv gRg^{-1},~ R)$. Choosing
gauge transformations as the right action, we get that constraints are
$\phi \equiv R=0$. So the ``gauge field" part in the action takes the form
\be
\int <R, g^{-1}dg> -(<\Lambda , R> + H(R,g))dt,
\ee
where $\Lambda \in {\cal A}$ is a Lagrange multiplier, and $H$ some
gauge invariant Hamiltonian.

Field theory generalization of (3.5) is the standard Yang-Mills theory.
In this section we consider corresponding generalization of (4.3)
with the action
\be
S=\int
{dt}~\left[ \int {d^{D-1}\vec x}
\left(\sum_{k=1}^{D-1}<R_k , g_k^{-1}\dot g_k> + e<A_0 , \phi>\right)
 - H \right]
\ee
where $g_k (\vec x,t)\in G$ and $R_k (\vec x ,t),
A_0 (\vec x ,t)\in {\cal A}$; $H$ is gauge invariant
Hamiltonian, $A_0$ --- Lagrange multipliers,
$\phi(\vec x,t)\equiv e\sum_{k=1}^{D-1}R_k(\vec x,t)$ --- constraints,
$e$ --- coupling constant (see below).

The ``1-form" $\sum_{k=1}^{D-1}<R_k , g_k^{-1}dg_k>$
defines the equal time Poisson brackets (see e.g.\ [16]):
\bea
\{ R_{k,a} (\vec x), R_{l,b} (\vec y) \}& =& \delta _{kl}\delta (\vec x -\vec
y)
f_{ab}^c R_{k,c} (\vec x) \nn \\
\{ g_k (\vec x), R_{l,a} (\vec y) \}& =& \delta _{kl}\delta (\vec x -\vec y)
(g_k T_a (\vec x)) \nn \\
\{ g_k (\vec x),g_l(\vec y) \}& =& 0
\eea
where the set $\{ T_a |~ T_a\in {\cal A} \}$ forms any basis of Lie algebra,
$ R_a \equiv <T_a , R>$ and the last two relations  are matrix equalities [16].
So for the constraints $\phi_a \equiv <T_a , \phi >$ we have
\be
\{ \phi _a (\vec x), \phi_b (\vec y) \} = \delta (\vec x -\vec y)
f_{ab}^c \phi _c (\vec x).
\ee
Corresponding gauge transformations are
\be
g_k \longrightarrow g_k \omega,~~~~~R_k \longrightarrow \omega^{-1}R_k \omega
\ee
and one can easily construct {\bf GIV}s such as
\be
g_{kl} =g_k g_l^{-1} \quad\mbox{and}\quad L_k = g_k R_k g_k^{-1}.
\ee
The Hamiltonian $H$ in (4.4) is any functional of such {\bf GIV}s.

Since (4.8) gives sufficient number of {\bf GIV}s we can use the scheme
described in Section 2. The first non-trivial case is the 3-dimensional
space-time. If we introduce $g = g_1g_2^{-1}$ as the $\xi^\mu$
variables and $R_1, R_2$ and $g_2$ as the $\eta $ variables of the scheme
(see (2.11)), then for the ``1-form" $\theta = <R_1,g_1^{-1}dg_1> +
<R_2,g_2^{-1}dg_2>$ (integration over ${\bf R}^2$ is assumed)
we immediately get
$$
<R_1+R_2,g_2^{-1}dg_2> + <g_2R_1g_2^{-1},g^{-1}dg>
$$
and after reduction we have
\be
\theta |_{\cal M}= <r,g^{-1}dg>,
\ee
where $r=g_2R_1g_2^{-1}$ is also {\bf GIV}.

So the structure of the 1-form is the same, only the number of variables
was reduced. One can check that this is true for other dimensions too.

It is clear that the phase spaces of the systems with 1-forms (4.1)
and (4.2) are essentially different and they cannot
be transformed to each other. But in field theory when one has the infinite
number of such spaces
there is a non-local transformation
 (see [17c]):
\be
A_k = \frac{1}{e}g_k^{-1}\partial_k g_k,~~~~~~~
E_k = - e g_k^{-1}\partial_k^{-1} (L_k) g_k,
\ee
which transforms the system (4.4)
into the Yang-Mills theory with the same gauge group $G$. Indeed, from (4.4)
and (4.10) one can check that
$$
\phi = \sum_{k=1}^{D-1}\partial_k E_k + e [A_k , E_k ]~~~~  \mbox {(Gauss law)}
$$
and
\be
 <E_k , \dot A_k> = <R_k , g_k^{-1}\dot g_k> +~~ \mbox {(total derivatives)}
\ee
To get the corresponding Hamiltonian of the Yang-Mills theory [17]:
$$
H=\frac{1}{2}\int d^{D-1}{\vec x} (\sum_{k=1}^{D-1}<E_k , E_k> +
\frac{1}{2}\sum_{k,l=1}^{D-1}<F_{kl} ,F_{kl}>)
$$
with $ F_{kl} = \partial_k A_l - \partial_l A_k +  e [A_k , A_l ] $,
one has to choose in (4.4)
\be
H = \frac{1}{2}\int {d^{D-1} {[e^2 <\partial_k^{-1} L_k , \partial_k^{-1} L_k>
+\frac{1}{e^2}<\partial_k (g_{kl}\partial_l g_{lk}) ,
\partial_k (g_{kl}\partial_l g_{lk}) >]}}.
\ee

So one can assume that the system (4.4) with the Hamiltonian (4.12) is
equivalent to the ordinary Yang-Mills theory with some boundary conditions
(which allow to invert (4.10) and neglect the total derivatives in (4.11).

Boundary behaviour is a subtle problem even for simple models of field
theory (see e.g.\ the Appendix). For the Yang-Mills theory it is too
complicated
and we do not consider it here.

Unfortunately the complicated form of the Hamiltonian (4.12) does not simplify
after the reduction procedure. For example, for the considered 3-dimensional
case the reduced Hamiltonian takes the form
\bea
H = \frac{1}{2}\int {d^2x} \Bigl[e^2 <\partial_1^{-1} r , \partial_1^{-1} r>
+ e^2 <\partial_2^{-1} l , \partial_2^{-1} l> + \nonumber \\
\frac{1}{e^2}
<\partial_1 (g\partial_2 g^{-1}) ,
\partial_1 (g\partial_2 g^{-1}>\Bigr],
\eea
where $l=grg^{-1}$ .

Gribov's ambiguity problem has stimulated many papers on the
gauge invariant description of
the Yang-Mills theory and the reduced system  (4.9),(4.13) is the one possible
version (literature and new results on this problem see in [1e]).
 The main problem of such approaches is a complicated form of the
Poincare generators in {\bf GIV}s [17]. For example, the Hamiltonian (4.13) is
non-local in fields an non-analytical in coupling constant. So the standard
perturbative quantization is not applicable.

Note that such Hamiltonian with corresponding symplectic form was
obtained in [17c] by Dirac's brackets formalism.

\vspace{1.cm}

    \section{Conclusions}

\vspace{1.cm}

  Of course, there was an essential progress
 in the study of constraint systems  since the
 paper [18], but from the point of view of
 practical applications still there is no universal approach. The method
 presented in this paper is a one possible
practical procedure towards the quantization
 of gauge theories.

 As it was mentioned in the introduction, for the gauge invariant
 systems there is an alternative way of quantization when one
``first quantizes and then reduces".  In general, there are two problems
in such approach:

 a) construction of physical states $|\Psi_{phys}>$ as the solutions of
the equations $\hat\phi_a|\Psi_{phys}>=0$,
where $\hat\phi$ are constraint operators.

 b) problem of scalar product for the physical states.

 Sometimes the first problem is only a technical one (for the Yang-Mills theory
 see [18]), but in general both this problems are related
 and need further investigation [20].

 In this paper we have not mentioned  other important methods such
 as the path integral approach [1b], [3] and BRST quantization [21]
 (for a review see [10a]).

 Quantization procedure is not unique even for the ordinary,
 non-constrained systems [8],[22]. It depends on the choice of
 canonical variables (if they globally  exist), operator ordering,
 etc. Therefore it is not surprising that different quantization
 procedures of constraint systems generally lead to the
 non-equivalent quantum systems [3],[23].

 As it was mentioned in Section 2, for a reduced classical system
generally
there are no global canonical coordinates and usual canonical
 quantization is not applicable. This, together with technical
 problems of classical reduction, was the main obstacle in general
 formulation of the approach ``first reduce and then quantize".

 Geometric quantization [9] and other ``new"
 quantization schemes  [7],[10a],[24] allow to quantize
 Hamiltonian systems without global canonical structure too. At the same time
 essential progress was made in the construction of classical reduction
 schemes [6]. Therefore for the wide class of constrained systems
 the quantization method ``first reduce and then quantize"
 seems to us to be technically preferable. Here it should be mentioned
 about the possible combination of the two
  quantization schemes: if a reduced classical system is complicated, then
  on the cotangent bundle of a reduced phase space one can
  construct new extended system with simple constraints and next use
the first way of  quantization [7],[10a,b].
 Of course, the question, which is
 the ``correct" quantum description of a given classical system, remains open.

\vspace*{0.5cm}

{\bf{Acknowledgments}\/}\\

\vspace*{0.5cm}

One of us (G.J.) is most grateful to I. Gogilidze, R. Loll, L. Lusanna,
L.V. Prokhorov, V. Rubakov and the colleagues from the seminar of Steklov
Mathematical Institute for helpful discussions.

The work has partial support under the INTS-93-1630 project
and JSPS grant.

\setcounter{equation}{0}
\def\theequation{A.\arabic{equation}}

\vspace*{1cm}

\begin{center}{\Large\bf Appendix}
\end{center}

The $2+1$ dimensional massive photodynamics is described by the
Lagrangian (see e.g.\ [25]):
 \be
 {\cal L} = -\frac{1}{4}F_{\mu \nu}F^{\mu \nu} -
 \frac{m}{4}\epsilon^{\mu \nu \sigma}
 F_{\mu \nu}A_\sigma .
 \ee
 We choose $g_{\mu\nu}=diag(+,-,-)$, $\epsilon^{012}=1 $ and
in the first order formalism [6] obtain
\be
S= \int dt \int_{R^2} d^2 x [(E_i -\frac{m}{2}\epsilon_{ij}A_j )\dot A_i
-\frac{1}{2}(E_i E_i + B^2 )+A_0 (\partial_i E_i - mB)],
\ee
where
$$
E_i\equiv F_{0i} \equiv \dot A_i - \partial_i A_0 ,~~~~
B \equiv \frac{1}{2}\epsilon_{ij}F_{ij}~~~~(\epsilon_{ij}\equiv\epsilon^{0ij})
$$
and we have neglected the boundary term
$\int_{R^2} d^2x \partial_i[A_0 (\frac{m}{2}\epsilon_{ij}A_j-E_i)]$.

 If we use ``1-forms" instead of time derivatives (see comment after (3.1)),
then the action (A.2) takes the form (2.1) with $A_0$ playing the role of
Lagrange multiplier.

  For
the reduction we choose $E_1$ and $E_2$  as the
 variables $\xi^\mu$,  $A_1$ as the additional
variable $\eta$ (see (2.11)) and  get
\be
\tilde S= \int dt \int_{R^2} d^2 x \left[\frac{1}{m}E_2\dot E_1 -
\frac{1}{2}[E_i E_i + \frac{1}{m^2}
(\partial_k E_k)^2] + \frac {d}{dt}\Theta\right],
\ee
where
$$ \Theta = \frac{1}{2}[E_1 A_1 +E_2 \hat K(A_1+\frac {1}{m}E_2)] $$
and the operator $\hat K \equiv \partial_1^{-1}\partial_2$ is assumed to be
symmetric due to the corresponding boundary conditions.

Neglecting the $\Theta$ term as the total derivative,
we get the local Hamiltonian theory with the
canonical commutation relations
\be
\{E_2 (x), E_1 (y)\} = m\delta^{(2)} (x-y)
\ee
and the quadratic Hamiltonian
\be
\frac{1}{2}\int_{R^2} d^2x [E_i E_i + \frac{1}{m^2}
(\partial_k E_k)^2].
\ee
The energy-momentum tensor also can be expressed only through
$E_1$ and $E_2$:
\be
T_{00}=\frac{1}{2}[E_i E_i + \frac{1}{m^2}(\partial_k E_k )^2 ],~~~~~
T_{0i}=\frac{1}{m}\epsilon_{ij}E_j (\partial_k E_k).
\ee

Let us briefly stop on the boundary conditions.
We can assume that
a boundary behaviour of the physical variables ($E_1, E_2$) should provide
the Poincare
invariance of the reduced system (A.3)-(A.6), while
a boundary behaviour of the fields of the initial system (A.1)
should allow the outlined reduction procedure.

Generators of the Poincare group (constructed from the
energy-momentum tensor (A.6)) generate transformations of $E_1$ and
$E_2$ according to the Poisson brackets (A.4). The class of functions
 $E_1(x)$ and $E_2(x)$ should be invariant under these transformations.
It is  natural to choose the class of smooth, rapidly vanishing at the
infinity functions.

For the diagonalization of the Hamiltonian and momentum let us take the Fourier
transformation:
\be
 E_j (x) = i\int d^2p \frac {e^{-i (p\cdot x)}}{2\pi}
\tilde  E_j (p)
\ee
and introduce the longitudinal and transverse components:
\be
\tilde E_j (p) = \frac {p_j}{|p|}e_1 (p) -
\frac{\epsilon_{jl}p_l}{|p|}e_2 (p),
\ee
 where $|p| = \sqrt {p_1^2+p_2^2}$.

 Then diagonalization
of the energy and momentum will be achieved in the variables
\bea
a(p)&=& \frac{\frac{\omega_ p}{m}e_1 (p) +
ie_2 (p)}{\sqrt {2\omega_ p}} e^{-i\varphi (p)} \nn \\
a^* (p)&=& \frac{\frac{\omega_ p}{m}e_1 (-p) -
ie_2 (-p)}{\sqrt {2\omega_p}} e^{i\varphi (p)},
\eea
with
$$\omega_p = \sqrt {|p|^2 + m^2}~~~{\mbox and}~~~
  e^{\pm i\varphi (p)}=\frac{p_1 \pm ip_2}{|p|}$$.

Note that for the chosen class of $E_1(x)$ and $E_2(x)$
the longitudinal and transverse components of $\tilde E_j (p)$ have
the singularity at the origin ($p=0$) and the phase factor $e^{i\varphi (p)}$
is necessary to cancel it.  On the other hand one can easily check
that the  class of smooth functions $a(p)$, $a^*(p)$ is Poincare invariant.
Just this phase factor was introduced in [25]
to avoid anomalies in the commutation relations of the Poincare
algebra of quantum operators.
As we have seen this phase factor is connected to the Poincare invariance of
the classical system too.

After description of the class of physical variables one can go back and
find the class of gauge potentials $A_\mu$. One can show that these
classes for
massive and ordinary photodynamics in (2+1) dimensions are different.

\end{document}